\documentclass[twocolumn,english,prl,groupedaddress]{revtex4}
\usepackage[T1]{fontenc}
\usepackage[latin9]{inputenc}
\usepackage{amsmath}
\usepackage{graphicx}
\usepackage{amssymb}
\usepackage{esint}

\makeatletter
\@ifundefined{textcolor}{}
{%
 \definecolor{BLACK}{gray}{0}
 \definecolor{WHITE}{gray}{1}
 \definecolor{RED}{rgb}{1,0,0}
 \definecolor{GREEN}{rgb}{0,1,0}
 \definecolor{BLUE}{rgb}{0,0,1}
 \definecolor{CYAN}{cmyk}{1,0,0,0}
 \definecolor{MAGENTA}{cmyk}{0,1,0,0}
 \definecolor{YELLOW}{cmyk}{0,0,1,0}
 }

\usepackage{bm}

\makeatother

\makeatother

\makeatother

\makeatother

\usepackage{babel}

\makeatother

\usepackage{babel}

\begin{document}

\title{Complex critical exponents for percolation transitions in Josephson-junction arrays, antiferromagnets, and interacting bosons}

\author{Rafael M. Fernandes and J\"org Schmalian}

\affiliation{Ames Laboratory and Department of Physics and Astronomy, Iowa State
University, Ames, IA 50011}

\date{\today }
\begin{abstract}
We show that the critical behavior of quantum systems undergoing a
percolation transition is dramatically affected by their topological
Berry phase $2\pi\rho$. For irrational $\rho$, we demonstrate that
the low-energy excitations of diluted Josephson-junctions arrays,
quantum antiferromagnets, and interacting bosons are spinless fermions
with fractal spectrum. As a result, critical properties not captured
by the usual Ginzburg-Landau-Wilson description of phase transitions
emerge, such as complex critical exponents, log-periodic oscillations
and dynamically broken scale-invariance. 
\end{abstract}


\maketitle
A fundamental aspect of the Ginzburg-Landau-Wilson (GLW) description
of phase transitions is scale invariance, which relies on the absence
of characteristic length and energy scales at criticality, leading
to the concept of universality \cite{Ma,Sachdevbook}. For instance,
near a quantum critical point (QCP), if a physical observable $O\left(T\right)$
transforms for an arbitrary scale transformation $b>0$ according
to $O\left(T\right)=b^{-x}O\left(b^{z}T\right)$, then we obtain a
power-law temperature dependence $O\left(T\right)\propto T^{x/z}$,
with universal critical exponent $x/z$. However, if scaling is valid
only for powers of a discrete value $b_{0}$, it follows that $O\left(T\right)=T^{x/z}Q\left(\ln T\right)$,
with $Q\left(t\right)$ a periodic function of period $z\ln b_{0}$.
Fourier expansion of $Q\left(t\right)$ yields: \begin{equation}
O\left(T\right)=\sum_{n=-\infty}^{\infty}\alpha_{n}T^{x/z+i2\pi n/\left(z\ln b_{0}\right)}\label{discrete_scale_inv}\end{equation}
with constant coefficients $\alpha_{n}$. Thus, the system is characterized
by a family of non-universal complex exponents. An invariant scale
$b_{0}$, leading to this \emph{discrete scale invariance,} is found
in several critical systems that either are out of equilibrium or
have an underlying built-in hierarchical structure (for a review,
see \cite{Sornette98}).

In this Letter, we show that complex critical exponents and log-periodic
behavior appear in a variety of disordered systems close to a percolation
QCP, such as Josephson-junction (JJ) arrays, quantum antiferromagnets
(QAF) and interacting bosons. Rather than being related to non-equilibrium
properties or to the fractality of the percolating cluster, in these
systems the invariant scale $b_{0}$ emerges naturally in their low-energy
excitation spectrum for certain values of their Berry phase $2\pi\rho$.

By calculating their specific heat and compressibility at the percolation
threshold, we show that, for rational $\rho$, large clusters have
the lowest excitation energies, giving rise to usual power-law behavior
below a crossover temperature $T^{*}$, which varies in a pronounced
non-monotonic way with respect to $\rho$ (see Fig. 1). For irrational
$\rho$, the low-energy properties are governed instead by degenerate
clusters of intermediate sizes, leading to the breakdown of continuous
scale invariance ($T^{*}\rightarrow0$). Remarkably, the sizes and
energies of these resonating clusters depend solely on the continued-fraction
expansion of $\rho$. As a result, when $\rho$ is a quadratic irrational,
the periodicity of its continued-fraction expansion gives rise to
an invariant scale $b_{0}$ and, consequently, to complex critical
exponents.

\begin{figure}
\begin{centering}
\includegraphics[width=1\columnwidth]{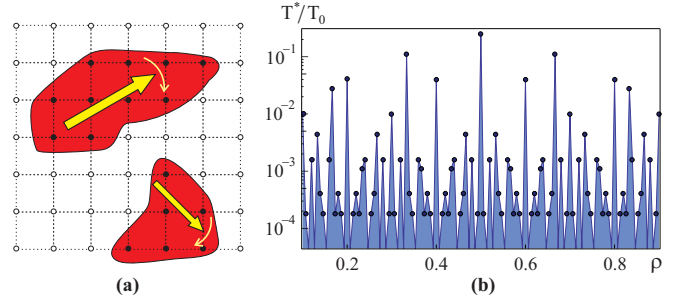} 
\par\end{centering}

\caption{(a) In the diluted quantum system, the global phase (arrow) of clusters
of connected grains (dark/red) coherently precesses due to the Berry
phase $2\pi\rho$. (b) Strong variation of the crossover temperature
$T^{*}$ below which scaling with real-valued exponents holds (blue
points, $T_{0}\sim U$). }

\end{figure}

To introduce a model for all the systems discussed above \cite{Sachdevbook},
consider an array of grains characterized by an XY order parameter
$\Psi_{j}=\left\vert \Psi_{0}\right\vert \exp\left(i\theta_{j}\right)$,
with phase $\theta_{j}$ and fixed amplitude $\left\vert \Psi_{0}\right\vert $.
The array is diluted on a regular lattice of dimension $d>1$, characterized
by a quenched random-site variable $\epsilon_{j}$ that takes the
values $0$ and $1$ with probabilities $P$ and $(1-P)$, respectively.
We consider the Hamiltonian \cite{Sachdevbook,Doniach,Fisher89}:

\begin{equation}
H=U\sum_{i}\epsilon_{i}\left(n_{i}-\rho\right)^{2}-\sum_{ij}\epsilon_{i}\epsilon_{j}J_{ij}\cos\left(\theta_{i}-\theta_{j}\right),\label{H}\end{equation}
where $n_{i}=-i\frac{\partial}{\partial\theta_{i}}$. $\rho$ can
be externally controlled and causes the phase to precess in time according
to $\partial\theta_{j}/\partial t=2U\rho$ (see Fig. 1). In JJ-arrays
\cite{Doniach,Fisher89}, $\Psi_{j}$ denotes the superconducting
order parameter, $J_{ij}$ is the Josephson coupling, $U$ is the
charging energy and $\rho$, related to the AC-Josephson effect, can
be changed by an external gate voltage. $\Psi_{j}$ can also represent
planar quantum rotors, associated with the low-energy modes of QAF
\cite{Sachdevbook}, where $\rho$ is proportional to a perpendicular
external magnetic field. For systems of interacting bosons\cite{Weichman08,Altman08},
which can be realized in optical experiments with cold atoms \cite{Bongs},
$2U\rho$ corresponds to the chemical potential $\mu$.

The effects of percolative dilution on all these systems have been
the subject of various experimental and numerical investigations \cite{Bongs,Yun06,Lv09,Vajk02,WangSandvik06}.
Here, we focus on the critical properties at the percolation threshold
$P_{c}$, where the density of clusters with $s$ connected occupied
sites varies as $N\left(s\right)\propto s^{-\tau}$, with $2<\tau\equiv d/D_{f}+1\leq2.5$
and $D_{f}$ the fractal dimension of the percolating cluster \cite{Stauffer}.
At low temperatures $T\ll\left\vert J_{ij}\right\vert $ and deep
in the ordered state of the clean system ($U<\left\vert J_{ij}\right\vert $),
the relative phase between grains inside each cluster is fixed, implying
that the entire cluster is characterized by a global phase \cite{VojtaSchmalian05B,Hoyos09,BrayAli}.
At $P_{c}$, contributions to the total specific heat of a single
cluster arise from the coherent precession of its global phase, $C$,
and from the excitations of internal collective (spin-wave) modes
that change the relative phase between grains, $C_{sw}$. As we will
show below, $C_{sw}$ is sub-leading; thus, similar to the behavior
in superparamagnets, each size-$s$ cluster can be treated effectively
as a big single rotor, with the corresponding action: \begin{equation}
\mathcal{A}_{s}=-\frac{s}{4U}\int_{0}^{\beta}d\bar{\tau}\left(\frac{\partial\theta\left(\bar{\tau}\right)}{\partial\bar{\tau}}-i\mu\right)^{2},\label{S}\end{equation}
which describes the coherent phase-precession due to both quantum
fluctuations and the Berry phase $2\pi\rho$. Notice that, unlike
the case of SU$(2)$ spins, the Berry phase of quantum rotors has
a topological character, since the imaginary part $\mathcal{A}_{\mathrm{Berry}}=is\rho\int_{0}^{\beta}d\bar{\tau}\frac{\partial\theta}{\partial\bar{\tau}}$
of $\mathcal{A}_{s}$ is independent on the time evolution of $\theta\left(\overline{\tau}\right)$,
enabling us to solve our problem using sums over winding numbers.
Shifting the imaginary time $\bar{\tau}\rightarrow\bar{\tau}/s$ in
(\ref{S}) eliminates the pre-factor $s$ at the expense of a cluster
size dependent temperature $T\rightarrow sT$ and, most importantly,
Berry phase $\rho\rightarrow s\rho$. This yields the free energy
scaling $F_{s}\left(\rho,T\right)=s^{-1}F_{1}\left(s\rho,sT\right)$,
from which we can derive scaling relations for the heat capacity $C_{s}\left(T\right)=-T\partial^{2}F_{s}/\partial T^{2}$
and the compressibility $\kappa_{s}\left(T\right)=-\partial^{2}F_{s}/\partial\mu^{2}$.
Here, the suffix $1$ ($s$) refers to quantities on a single site
(single cluster). Thus, macroscopic quantities can be calculated by
averaging over all clusters, i.e. $\mathcal{O}\left(\rho,T\right)=\sum_{s}N\left(s\right)\mathcal{O}_{s}\left(\rho,T\right)$.

Let us first revisit the results for $\rho=0$, where universal power-law
behavior was previously found \cite{VojtaSchmalian05B}. The low-temperature
specific heat of a cluster is given by $C_{s}\left(\rho=0\right)\propto\exp\left(-U/sT\right)$,
i.e. the typical excitation energy of a cluster decreases \textit{monotonically}
with its size, $\varepsilon_{s}=U/s$. Then, the low-energy behavior
is dominated by large clusters and we can replace the sum over $s$
by an integral, obtaining, for $T\ll U$, \begin{equation}
C\left(\rho=0,T\right)\propto T^{d/z_{r}},\label{Cbos}\end{equation}
 where the dynamic scaling exponent $z_{r}=D_{f}$ was introduced.
This result was also found in detailed computer simulations \cite{Sandvik02,Yu05}.
Consider now a finite Berry phase $\rho\neq0$. Solving the problem
of a single quantum rotor and using the scaling $T\rightarrow sT$,
$\rho\rightarrow s\rho$ derived from the action (\ref{S}), we find
the spectrum of a single cluster $E_{m}=U\left(m-s\rho\right)^{2}/s$,
with $m\in\mathbb{Z}$. Therefore, the lowest excitation energy is
(see also \cite{Weichman08,Altman08}): \begin{equation}
\varepsilon_{s}=\frac{U}{s}\left(1-2\left|\rho_{s}\right|\right),\label{excit}\end{equation}
where $\rho_{s}=s\rho-\left\lfloor s\rho+\frac{1}{2}\right\rfloor $
and $\left\lfloor x\right\rfloor $ is the integer part of $x$, i.e.
$\left\lfloor x+\frac{1}{2}\right\rfloor $ is the integer closest
to $x$, such that $\left|\rho_{s}\right|\leq1/2$. Note that $\rho_{s}$
depends on the droplet size in a highly non-monotonic way, reflecting
the periodicity of the Berry phase (see Fig. 2). Now, not only large
clusters yield small excitation energies, but also intermediate-size
clusters with $\left|\rho_{s}\right|\lesssim1/2$.

\begin{figure}
\begin{centering}
\includegraphics[width=0.9\columnwidth]{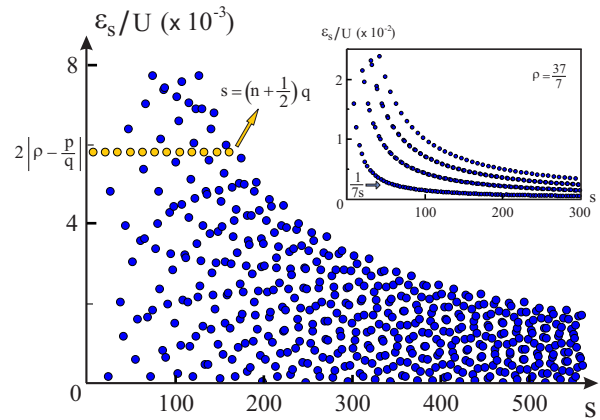} 
\par\end{centering}

\caption{Excitation energy $\varepsilon_{s}$ as function of the cluster size
$s$ for the irrational $\rho=\sqrt{7}$. The highlighted points correspond
to energetically degenerate clusters associated to the Diophantine
approximant $p/q=37/14$, and are responsible for a jump in the integrated
density of states. The inset shows the spectrum for the rational $\rho=37/7$,
characterized by four well-defined branches. }

\end{figure}

These low-energy excitations can be described by spinless fermions,
yielding the well-known fermionic expressions $C_{1}\left(\omega,T\right)=\left(\omega/T\right)^{2}n_{F}\left(\omega\right)n_{F}\left(-\omega\right)$
and $\kappa_{1}\left(\omega,T\right)=\left(1/T\right)n_{F}\left(\omega\right)n_{F}\left(-\omega\right)$
for the single site specific heat and compressibility, with $n_{F}\left(\omega\right)$
the Fermi function. In Fig. 3 we demonstrate this numerically by comparing
$C\left(T\right)$ obtained from the exact energy spectrum and from
the fermionic expression only. We can also show this result analytically:
Consider, for definiteness, $0\leq\rho\leq1/2$. The spectrum of a
single rotor, including degeneracies, can be generated from a model
of effective interacting fermions and bosons, with occupation numbers
$n_{f}=f^{\dagger}f$ and $n_{b}=b^{\dagger}b$, respectively \begin{equation}
H_{0}=U\left(n_{b}+\left(1-2\rho\right)n_{f}+\rho\right)^{2}.\label{supersymmetry}\end{equation}
 For $\rho=0$, one recovers the $N=2$ super-symmetric description
of a rotor \cite{Correa07}. Expanding for $\rho\simeq1/2$, we obtain
instead $H_{0}\simeq\varepsilon_{1}n_{f}+Un_{b}+Un_{b}^{2}+2\varepsilon_{1}n_{b}n_{f}$.
While the excitation energy of single bosons is $U$, the fermionic
excitation energy is $\varepsilon_{1}=U\left(1-2\rho\right)\ll U$.
Thus, at sufficiently low temperatures, bosons are diluted and the
interaction terms can be neglected, implying that free spinless fermions
are the dominant excitations of the system. The limit $\rho\rightarrow1/2$
also plays an important role in the excitation spectrum of droplets
in the Bose-glass phase \cite{Weichman08} and in the insulating phase
of interacting bosons in a disordered chain \cite{Altman08}.

Let us now consider $\rho=p/q$ to be rational, i.e. $p$ and $q$
are integers with no common divisors. Due to the periodicity of the
Berry phase term, the number of distinct values of $\rho_{s}$ is
of the order of $q/2$, defining well separated branches in $\varepsilon_{s}$,
all decaying as $s^{-1}$, with lowest branch $\varepsilon_{s}\simeq U\left(qs\right)^{-1}$
(see inset of Fig. 2). For $T\ll U/q^{2}$, the problem is virtually
the same as for $\rho=0$, leading to a heat capacity dominated by
very large clusters as given in Eq.\ref{Cbos} with the same exponent
$z_{r}$, as shown in Fig. 3. For the compressibility, we obtain $\kappa\left(T\right)\propto T^{d/z_{r}-1}$,
with a Wilson ratio $\left(\kappa T\right)/C\approx0.3$ for $d=2$.
Note, from Fig. 3, that the crossover temperature $T^{*}$ changes
as $q^{-2}$ and is insensitive to $p$. Thus, systems with similar
values of $\rho$ can have very different $T^{*}$ (see Fig. 1).

\begin{figure}
\begin{centering}
\includegraphics[width=0.9\columnwidth]{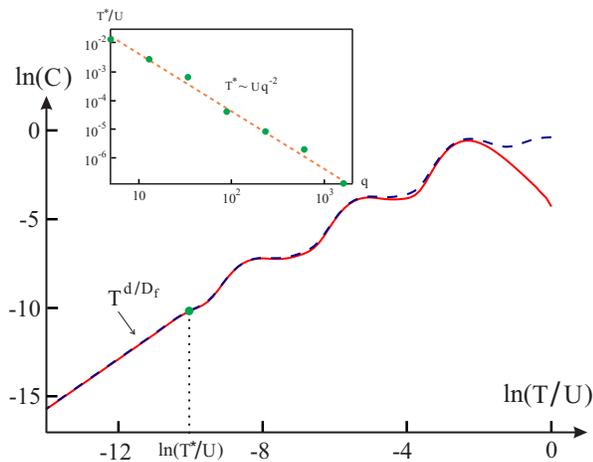} 
\par\end{centering}

\caption{Specific heat $C$ as function of temperature $T$ for the rational
$\rho=55/89$. The dashed blue line is the exact result and the solid
red line is the fermionic approximation. The onset of power-law behavior
is marked by $T^{*}$(dotted line). The inset shows the log-log variation
of $T^{*}$ with respect to the denominator $q$ for a series of rationals
with comparable values, $\rho=\left\{ \frac{3}{5},\frac{8}{13},\frac{21}{34},\frac{55}{89},\frac{144}{233},\frac{377}{610},\frac{987}{1597}\right\} $. }

\end{figure}

The natural question is: what happens in the regime $U/q^{2}\ll T\ll U$
for very large $q$? For irrational $\rho$, $T^{*}\rightarrow0$
and this regime prevails down to the lowest energies. Indeed, when
$\rho$ is irrational, the sequence $\rho_{s}$ is uniformly distributed
between $-1/2$ and $1/2$ \cite{Weyl1916}, i.e. there are finite-size
droplets with arbitrarily low excitation energy $\varepsilon_{s}$.
It is convenient to introduce the averaged integrated density of fermionic
states, $D\left(\omega\right)=\sum_{s}s^{-\tau}\theta\left(\omega-\varepsilon_{s}\right)$;
from the periodicity of $\varepsilon_{s}$ with respect to $\rho$
(i.e. summing over winding numbers), we obtain, for $\omega\ll U$:
\begin{equation}
D\left(\omega\right)=\zeta\left(\tau-1\right)\frac{\omega}{U}+\sum_{s,\lambda=\pm}\frac{f_{\mathrm{saw}}\left(sx_{\lambda}\right)}{s^{\tau}},\label{DOS}\end{equation}
 where $x_{\pm}=\omega/U\pm2\rho$, $\zeta\left(x\right)$ is the
zeta function and $f_{\mathrm{saw}}\left(x\right)$ is the sawtooth
function, which has period $2$ and unit jumps at every odd integer:
$f_{\mathrm{saw}}\left(x\right)=\left(-x+2n\right)/2$ for $2n-1<x<2n+1$.
These jumps give rise to discontinuities in $D\left(\omega\right)$
at frequencies:\begin{equation}
\omega_{j}=2U\left\vert \rho-\frac{p_{j}}{q_{j}}\right\vert ,\label{jump}\end{equation}
 where $p_{j}$ is an odd integer and $q_{j}$ is even. At $\omega\ll U$,
the fractions $p_{j}/q_{j}$ that satisfy Eq. (\ref{jump}) are the
ones that best approximate $\rho$, i.e. the Diophantine approximants,
which are given by the convergents of the continued fraction expansion
of $\rho$ \cite{Hardy}. Physically, the jumps are a consequence
of the existence of a set of energetically degenerate (i.e. {}``resonating'')
clusters with sizes that are odd multiples of $q_{j}/2$, $s=\left(n+1/2\right)q_{j}$
(see Fig. 2). Summing over all these clusters in Eq. (\ref{DOS}),
we find that each jump in $D\left(\omega\right)$ is given by $\Delta_{j}=q_{j}^{-\tau}\zeta\left(\tau\right)\left(2^{\tau}-1\right)$.

Back to Eq. \ref{DOS}, we find that the regular part of the sawtooth
function cancels out the linear in $\omega$ term. Thus, the frequency
dependence of $D\left(\omega\right)$ is governed by the successive
jumps $\Delta_{j}$ at $\omega_{j}$ of Eq. (\ref{jump}) and, consequently,
by the sequence of convergents of the continued fraction expansion
of $\rho$ with even denominator $q_{j}$. Although the determination
of this sequence for an arbitrary irrational $\rho$ is an outstanding
problem in number theory, it is simplified in the case of quadratic
irrationals, which have periodic continued fraction expansions \cite{Hardy}.
Then, one finds that the sequence of even $q_{j}$ is also periodic.

In fact, for quadratic irrationals with a single period $a\in\mathbb{Z}$,
which are solutions of the algebraic equation $y^{2}-ay-1=0$, we
find that if $q_{j}$ is even, so is $q_{j+N}$, with $N=2+\textrm{mod}\left(a,2\right)$.
Consequently, the distance between consecutive jumps is a constant
in log-scale, $\ln\left(\omega_{j}/\omega_{j+1}\right)=2N\ln y_{+}$,
as well as the ratio between their amplitudes, $\ln\left(\Delta_{j}/\Delta_{j+1}\right)=\tau N\ln y_{+}$,
where $y_{+}$ is the positive solution of the algebraic equation.
Using these properties, we can show that $D\left(\omega\right)$ is
a fractal function with fractal dimension $d_{\omega}=\tau/2$, characterized
by a power-law decay in $\omega$ and periodic oscillations in $\ln\omega$.
Since $C\left(T\right)=\int d\omega\:\nu\left(\omega\right)C_{1}\left(\omega,T\right)$,
where $\nu=dD/d\omega$ is the density of states, we obtain:

\begin{equation}
C\left(T\right)=T^{d/z_{ir}}A\left(\ln T\right),\label{C_irr}\end{equation}
 where $z_{ir}=2D_{f}d/\left(D_{f}+d\right)$, i.e. $z_{ir}>z_{r}$,
and $A\left(t\right)$ is a periodic function of period $z_{ir}\ln b_{0}\equiv2N\ln y_{+}$.
For the compressibility, we find $\kappa\left(T\right)=T^{\left(d/z_{ir}\right)-1}B\left(\ln T\right)$,
where $B\left(t\right)$ has the same period as $A\left(t\right)$.
Thus, the system has complex critical exponents as in Eq. (\ref{discrete_scale_inv}).
In Fig. 4, we show numerical results for $\rho=\sqrt{2}$; we also
verified numerically that the scaling form in Eq. \ref{C_irr} holds
for quadratic irrationals $\rho$ with more complicated continued-fraction
periods. For non-quadratic irrationals, our numerical calculations
indicate that Eq. \ref{C_irr} still describes the critical behavior,
but now $A\left(t\right)$ oscillates irregularly without a well-defined
period.

The breakdown of continuous scale invariance for irrational $\rho$
can be attributed to the resonating clusters with arbitrarily low
excitation energies, as they cause jumps in the entire spectrum, prohibiting
to replace $\sum_{s}\rightarrow\int ds$ in Eq. \ref{DOS}. For rational
$\rho$, such a replacement is allowed, leading to full scaling $s\rightarrow s/b^{D_{f}}$
and to Eq. \ref{Cbos}. Yet, when $\rho$ is a quadratic irrational,
the dynamically broken scale invariance is partially restored as discrete
scale invariance. In this case, the periodic structure of the continued-fraction
expansion of $\rho$ gives rise to log-periodic relations between
sizes $s=\left(n+1/2\right)q_{j}$ and energies $\omega_{j}=2U\left\vert \rho-p_{j}/q_{j}\right\vert $
of different sets of resonating clusters, establishing an invariant
scale $b_{0}$ and complex critical exponents $\left(d/z_{irr}\right)+in\pi/\left(N\ln y_{+}\right)$
for $C\left(T\right)$.

\begin{figure}
\begin{centering}
\includegraphics[width=0.9\columnwidth]{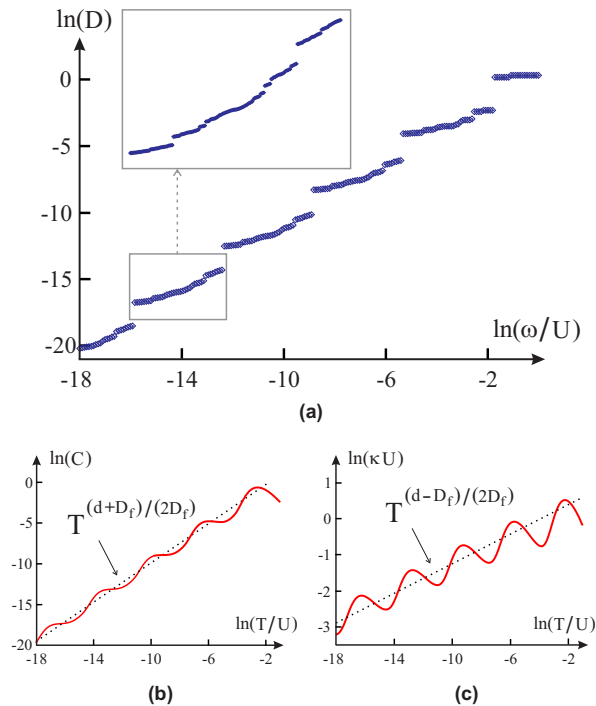} 
\par\end{centering}

\caption{(\textbf{a}) Frequency dependent integrated density of states $D\left(\omega\right)$,
as well as the temperature dependent (\textbf{b}) specific heat $C\left(T\right)$
and (\textbf{c}) compressibility $\kappa\left(T\right)$ for the system
with $\rho=\sqrt{2}$. Dashed lines show the underlying power-law
behavior superimposed to the log-periodic oscillations. Inset is a
zoom of $D\left(\omega\right)$.}

\end{figure}

Going back to the contribution of the spin-waves, their density of
states is $\nu_{sw}\propto k^{d_{s}-1}$ at $P_{c}$, where the fracton
dimension $d_{s}$ characterizes the spectrum of the eigenvalues $k^{2}$
of the Laplacian on the percolating cluster \cite{Stauffer,BrayAli}.
From the spin-waves dispersion $\Omega_{s,k}^{2}=\varepsilon_{s}^{2}+c^{2}k^{2}$
and the scaling properties of the fermionic density of states, we
find $\mathcal{\mathcal{O}}_{sw}\propto\mathcal{O}T^{\phi}$ for both
$\mathcal{O}=C,\kappa$, with $\phi=d_{s}-1$ ($\phi=d_{s}-1/2$)
for rational (irrational) $\rho$. As $d_{s}>1$ \cite{Stauffer},
it follows that $\phi>0$. Since the internal modes are sub-leading
compared to the coherent modes, the spectrum of a single cluster depends
solely on its size and not on its shape, in accordance to Eq. \ref{S}.

In summary, we solved the $XY$ quantum-rotor problem at low $T$
and close to the percolation threshold, which describes diluted systems
as diverse as JJ-arrays with d.c.-bias voltage, canted QAF in a perpendicular
magnetic field, and interacting bosons coupled to a particle reservoir.
Their topological Berry phase $2\pi\rho$ dramatically alters the
percolation QCP, since the low-$T$ behavior is governed by emergent
spinless fermions with fractal spectrum, giving rise to generally
irregular $\log T$-oscillations of thermodynamic variables. While
for irrational $\rho$ they persist to $T\rightarrow0$, for rational
$\rho=p/q$ they occur in the temperature range $q^{-2}\lesssim T/U\lesssim1$,
which can be broad for $q\gg1$. Remarkably, for a quadratic irrational
$\rho$, they become regular, leading to complex critical exponents.
Our results demonstrate that the quantum criticality in disordered
systems governed by a topological Berry phase is beyond the GLW paradigm
of critical systems.

We thank M. Axenovich for pointing out Ref.\cite{Weyl1916} to us.
This research was supported by the U.S. DOE, Office of BES, Materials
Sciences and Engineering Division.

\end{document}